\documentclass[fleqn,usenatbib]{mnras}

\usepackage{newtxtext,newtxmath}
 
\usepackage[T1]{fontenc}

\DeclareRobustCommand{\VAN}[3]{#2}
\let\VANthebibliography\thebibliography
\def\thebibliography{\DeclareRobustCommand{\VAN}[3]{##3}\VANthebibliography}

\usepackage{graphicx}	
\usepackage{amsmath}	
\usepackage{amssymb}	
\usepackage{xcolor}

\title[Combining multiple RSFs]{Strengthening nuclear symmetry energy constraints using multiple resonant shattering flares of neutron stars with realistic mass uncertainties}

\author[D. Neill et al.]{
Duncan Neill$^{1}$\thanks{E-mail: dn431@bath.ac.uk},
David Tsang$^{1}$,
William G. Newton$^{2}$
\\
$^{1}$Department of Physics, University of Bath, Claverton Down, Bath, BA1 1AL\\
$^{2}$Department of Physics and Astronomy, Texas A\&M University-Commerce, Commerce, TX, 75429-3011
}

\date{Accepted XXX. Received YYY; in original form ZZZ}

\pubyear{2015}

\begin{document}
\label{firstpage}
\pagerange{\pageref{firstpage}--\pageref{lastpage}}
\maketitle

\begin{abstract}
With current and planned gravitational-wave (GW) observing runs, coincident multimessenger timing of Resonant Shattering Flares (RSFs) and GWs may soon allow for neutron star (NS) asteroseismology to be used to constrain the nuclear symmetry energy, an important property of fundamental nuclear physics that influences the composition and equation of state of NSs. In this work we examine the effects of combining multiple RSF detections on these symmetry energy constraints, and consider how realistic uncertainties in the masses of the progenitor NSs may weaken them.
We show that the detection of subsequent multimessenger events has the potential to substantially improve constraints beyond those obtained from the first, and that this improvement is insensitive to the mass of the NSs which produce the RSFs and its uncertainty. This sets these asteroseismic constraints apart from bulk NS properties such as radius, for which the NS mass is highly important, meaning that any multimessenger RSF and GW events can equally improve our knowledge of fundamental physics.
\end{abstract}

\begin{keywords}
dense matter -- stars: neutron -- stars: oscillations -- neutron star mergers -- gravitational waves -- equation of state
\end{keywords}

\section{Introduction}

Neutron stars (NSs) provide a unique environment in which to study fundamental physics, as they consist of matter that is both extremely dense and neutron-rich without having the high temperatures required to produce these conditions in terrestrial experiments. There is a well established link between the composition and structure of NSs and the zero temperature equation of state (EOS) of bulk nuclear matter \citep{Lattimer_2001,LiKrastev2019}, which describes the binding energy of nuclear matter as a function of density and asymmetry (proton-to-neutron ratio). In particular, the neutron-rich matter found in NSs is sensitive to the asymmetry dependence of the nuclear EOS, which is encapsulated by the nuclear symmetry energy: the difference in binding energy between symmetric (N=Z) and pure neutron (N=A) matter. The symmetry energy is also important for various properties of nuclei, such as the thickness of neutron skins (the neutron-rich layer that surrounds the more symmetric core of a neutron-rich nucleus)~\citep{chen2010density,piekarewicz2012electric} and certain giant resonances (which excite collective oscillations of nucleons within nuclei, some of which produce relative displacements between the protons and neutrons)~\citep{trippa2008giant,RocaMaza2013Giant}. Of particular interest are recent precision measurements of the neutral weak form factors of $^{48}$Ca and $^{208}$Pb \citep{Adhikari2021PREXII,Adhikari2022CREX}, from which the neutron skin can be extracted, which have been shown to in tension with each other for the majority of nuclear energy-density functionals \citep{Reed_CREX2022,Zhang_CREX2022,Reinhard_CREX2022,Yuksel_CREX2022}.  Furthering our understanding of the symmetry energy using probes independent from current nuclear experimental techniques is important to resolve these tensions and learn more about fundamental physics and improve our ability to model dense matter.

Various methods exist to infer structural or compositional properties of NSs from astronomical observations \citep{Lattimer2007Prognosis,raithel2019constraints}, which can subsequently be used to impose constraints on the nuclear symmetry energy through NS modelling. Of particular interest are the transient gravitational waves (GWs) detected from Neutron Star-Neutron Star (NSNS)~\citep{abbott2017gw170817,abbott2020gw190425} and Black Hole-Neutron Star (BHNS)~\citep{abbott2021observation} binary mergers, which have opened the way to a new era of multimessenger NS astronomy involving GWs and electromagnetic counterparts~\citep[e.g.][]{Smartt2017kilonova,margalit2017constraining,radice2018GW170817,troja2018,abbott2018gw170817}. While at this time only a small number of such transients have been detected, the O4 LIGO GW observing run has already detected multiple BHNS merger candidates, and with the increased sensitivity of LIGO and Virgo in this and future runs we can expect to observe many more mergers involving NSs in the near future.

Other methods to measure NS properties include pulse-profile modelling of X-ray emission~\citep{riley2019NICER,Miller2019NICER,riley2021NICER,Miller2021NICER}, measuring the relativistic Shapiro delay of radiation from pulsars in binary systems~\citep{Fonseca2021Refined}, and modelling outbursts from low-mass X-ray binaries~\citep{Galloway2008Thermonuclear,Ozel2016Dense,Nattila2017Neutron}. However, a common theme of these methods is that they constrain NS properties which are sensitive to the nature of matter within the NS core \citep{Essick_2021}, where most of the star's mass is located. Matter within the ultra-dense NS core may transition to an exotic phase, but the type of matter it becomes and the density at which the transition occurs are highly uncertain, meaning that while constraints on core-dependent properties are useful for testing various exotic models they can not reliably be used to constrain properties of nucleonic matter, such as the symmetry energy. Instead, we must look for ways to probe the lower density -- but still very dense and neutron-rich -- NS crust, were matter can more confidently be assumed to be nucleonic. 

Perhaps the most compelling methods to probe specific properties of matter within NSs are those that involve asteroseismic normal modes. Different families of modes are sensitive to different NS properties~\citep{mcdermott1988nonradial}, some of which are dominated by the composition and structure of the NS crust. Various NS phenomena that may contain signatures of modes have been identified~\citep{lai1994resonant,Kokkotas1995Tidal,Andersson1998Unstable,Duncan1998Global,Israel2005Discovery,Strohmayer2005Discovery,Levin2007theory,tsang2012resonant,Sotani2024Shear}, which could allow for features of these modes and the NS properties on which they depend to be constrained.

\citet{tsang2012resonant} proposed the existence of resonant shattering flares, and that coincident detection of such a flare alongside gravitational waves could allow a measurement of the frequency at which the crust-core interface mode oscillates. Resonant shattering flares (RSFs) are brief flares of gamma-rays produced when a NS normal modes is resonantly excited by the tidal field of its inspiralling binary partner and reaches an amplitude sufficient to deform the NS crust beyond its breaking strain~\citep{tsang2012resonant}. This causes the crust to repeatedly fracture while the mode is near resonance, resulting in a build-up of seismic energy that eventually causes the crust to shatter upon reaching its elastic limit, scattering the seismic waves to higher frequencies. A strongly magnetised NS can then have its magnetic field couple to these high frequency waves, resulting in the emission of a fireball. During a mode's resonance multiple such fireballs can be emitted~\citep{neill2022resonant}, and collisions between them will produce internal shocks capable of scattering electrons to high energies, resulting in non-thermal synchrotron emission detectable as a brief flare of gamma-rays: a RSF. Multimessenger detection of GWs and a RSF from an inspiralling binary would allow us to measure the frequency of the resonant mode, as the GW frequency directly informs us of the orbital frequency, which at the time of the RSF will be around half the frequency of the (quadrupolar) resonantly excited mode.

As a mode must be strongly excited in order to shatter the NS crust, resonant shattering flares can only be triggered by modes that become resonant shortly before a NSNS or BHNS merger, when the tidal forces between a NS and its binary partner are strong. Additionally, the mode must primarily displace matter within the NS crust in order to reach its breaking strain efficiently. In previous works the crust-core interface mode ($i$-mode) was identified to satisfy these requirements~\citep{tsang2012resonant} and its frequency was found to be sensitive to the shear speed at the base of the NS crust~\citep{neill2021resonant}. The shear speed is sensitive to the proton fraction of the crust and thus in turn depends on the nuclear symmetry energy. A measurement of the $i$-mode's frequency obtained from multimessenger GW and RSF timing could therefore be used to constrain the symmetry energy at around half nuclear saturation density.

\citet{Neill2023Constraining} examined these symmetry energy constraints by performing Bayesian inference of the parameters of a NS meta-model using injected $i$-mode frequency data, and found that a single multimessenger event would provide constraints similar in strength to those from terrestrial nuclear experiments. However, that work fixed the mass of the NS from which the $i$-mode frequency was measured, ignoring any consequences of mass uncertainty. Given that other prominent NS observables -- such as radius or tidal deformability -- can have strong dependences on the NS's mass, we wish to examine the mass-dependence of the $i$-mode frequency, to determine its effect on the nuclear constraints obtainable from RSFs. It was also suggested in~\citet{neill2022resonant} that RSFs might not be uncommon, in which case we may be able to obtain multiple measurements of the $i$-mode frequency in the near future.The first aim of this work is therefore to investigate whether accounting for realistic uncertainty in the mass of a NS (inferred from the GW signal of its binary merger) significantly affects the symmetry energy constraints obtained from a measurement of its $i$-mode frequency. The second is to determine whether those symmetry energy constraints could be improved by combining data from multiple multimessenger GW and RSF detections, as subsequent events will reduce the effect of statistical uncertainties in the data and allow us to probe the mass-frequency relationship.

We will begin in Section~\ref{sec:f-vs-m} by briefly describing the NS meta-model we use to generate NS models from a set of nuclear physics parameters, and then will examine the relationship between NS mass and $i$-mode frequency. After that, in Section~\ref{sec:population} we will construct an artificial set of merging NSNS and BHNS binaries and use standard GW analysis methods to obtain realistic uncertainties for the masses of the objects in these binaries.These masses and their uncertainties will then be used in Section~\ref{sec:RSF_inference} along with injected $i$-mode frequency measurements -- which include estimates of the uncertainty resulting from coincident RSF and GW timing -- to investigate how combining multiple multimessenger RSF and GW events could affect symmetry energy constraints. In Section~\ref{sec:discuss} we will discuss the results and what they mean for how $i$-mode frequency measurements could be used in the future, as well as some of the limitations of this work. Finally in Section~\ref{sec:conclude} we will give our conclusions.

\section{Modelling neutron stars}\label{sec:f-vs-m}
\subsection{Neutron star meta-model and choice of injected model}\label{sec:model}
Our method for generating NS models from a set of nuclear physics parameters has been described in several previous works~\citep{Newton:2013aa,balliet2021prior,neill2021resonant}, so here we shall only give a brief overview. The binding energy of bulk nuclear matter can be expanded in isospin asymmetry ($\delta=1-2x_{\rm p}$, with $x_{\rm p}$ being the proton fraction) to give
\begin{align}
    E(n,\delta)=E_0(n) + \delta^2 E_{\rm sym} + \dots,
\end{align}
\noindent with the symmetry energy $E_{\rm sym}$ being the energy required to deviate from an equal concentration of neutrons and protons. The symmetry energy can further be expanded in density around nuclear saturation ($n_{\rm sat}\sim0.16\text{ fm}^{-3}$) with expansion parameters ($J$,$L$,$K_{\rm sym}$,$Q_{\rm sym}$,$\ldots$). We model the strong force between nucleons in bulk matter using a Skyrme type phenomenological effective potential with an extended density dependence~\citep{Skyrme1958effective,Davesne2016,Zhang2016_ExtSkyrme, Lim:2017aa}. The phenomenological parameters of this model can be related to the symmetry energy parameters, with the extended density dependence allowing three symmetry energy parameters ($J$, $L$ and $K_{\rm sym}$) to be chosen independently, rather than just two which is the case in most Skyrme models \citep{Dutra:2012wd}. In the NS inner crust this model's energy density functional is used with the compressible liquid droplet model (CLDM) to obtain the equilibrium structure of spherical nuclear clusters and pure neutron fluid \citep{Newton:2013aa,balliet2021prior,Newton:2021tg}, while in the core it is used to obtain the composition of uniform matter. The transition between these regions is located where the energy density of uniform matter matches the energy density obtained from the CLDM. 
This meta-model produces NS crusts and cores that are connected in a consistent way, allowing for simultaneous systematic exploration of how core and crust-dependent neutron star observables inform us of nucleonic matter properties.

As the Skyrme model describes nucleonic matter, it might not be valid at the very high densities found within the cores of massive NSs, where exotic phases of matter may appear. To avoid assumptions about the nature of matter in NS cores we therefore switch to a polytropic NS EOS model at $1.5n_{\rm sat}$, with a piecewise transition at $2.7n_{\rm sat}$ to allow more freedom in the core model \citep{Read2009Constraints,Steiner2010equation}. This gives us two additional parameters: the polytropic indices $\gamma_1$ and $\gamma_2$. These polytropes gives us sufficient model freedom explore a wide range of NS cores, and the restrictions they place on the core model and the lack of information they provide about its composition should not be important in this work, as the $i$-mode is insensitive to the core. While they do not have the freedom to explore in detail the range of possible cores introduced when allowing for exotic matter, they do reproduce the consequence of exotic matter that is most significant for this work, which is that properties of matter within the NS core will not be directly dependent on the symmetry energy parameters.

The parameters of this meta-model are therefore $J$, $L$, $K_{\rm sym}$, $\gamma_1$ and $\gamma_2$. For any set of values for these parameters we can construct a unique NS EOS including relevant compositional properties of the crust, such as the shear modulus. Such an EOS can then be used in the TOV equations \citep{oppenheimer1939massive,tolman1939static} to obtain equilibrium NS structures, which can in turn be used with \citet{yoshida2002nonradial}'s linearised relativistic pulsation equations to calculate (in the Cowling approximation) the frequency of the quadrupolar $i$-mode ($f_{2i}$). The TOV equations can also be augmented to calculate NS tidal deformability \citep{hinderer2010tidal} (the degree to which a NS is deformed by an external tidal field), which is relevant for GWs.

For the purpose of this work we choose a set of parameter values to inject, which we will attempt to recover using $i$-mode frequency measurements. For the symmetry energy parameters we choose values that are generally consistent with experimental constraints: $J=31 \text{ MeV}$, $L=50 \text{ MeV}$ and $K_{\rm sym}=-100 \text{ MeV}$. For the core polytrope parameters meanwhile we choose $\gamma_1=\gamma_2=3.0$, as for our chosen symmetry energy parameter values they result in a mass-radius relationship (shown in Figure~\ref{fig:mfrt_vary1}) which broadly agrees with astrophysical constraints~\citep{Ozel2016Dense,abbott2018gw170817,riley2019NICER,Miller2019NICER,riley2021NICER,Fonseca2021Refined}, with $R_{1.4}=12.2 \text{ km}$ and $M_{\rm max} = 2.17 \text{ M}_{\odot}$.

\subsection{Relationship between NS mass and i-mode frequency}
\begin{figure}
\centering
\includegraphics[width=0.45\textwidth,angle=0]{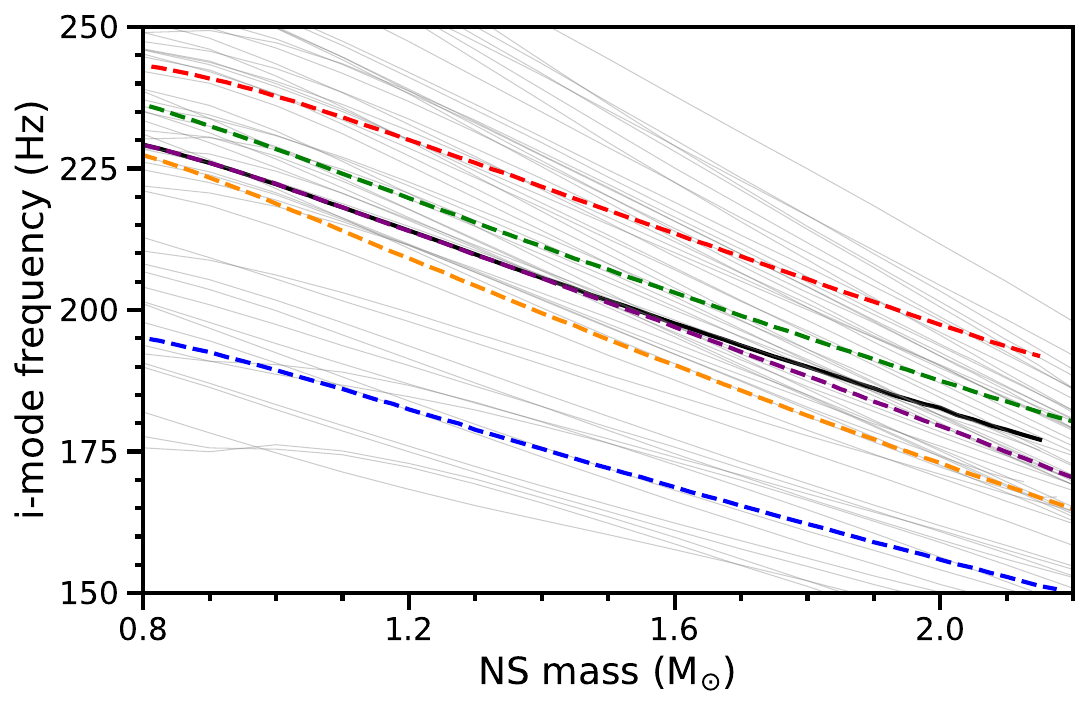}
\includegraphics[width=0.45\textwidth,angle=0]{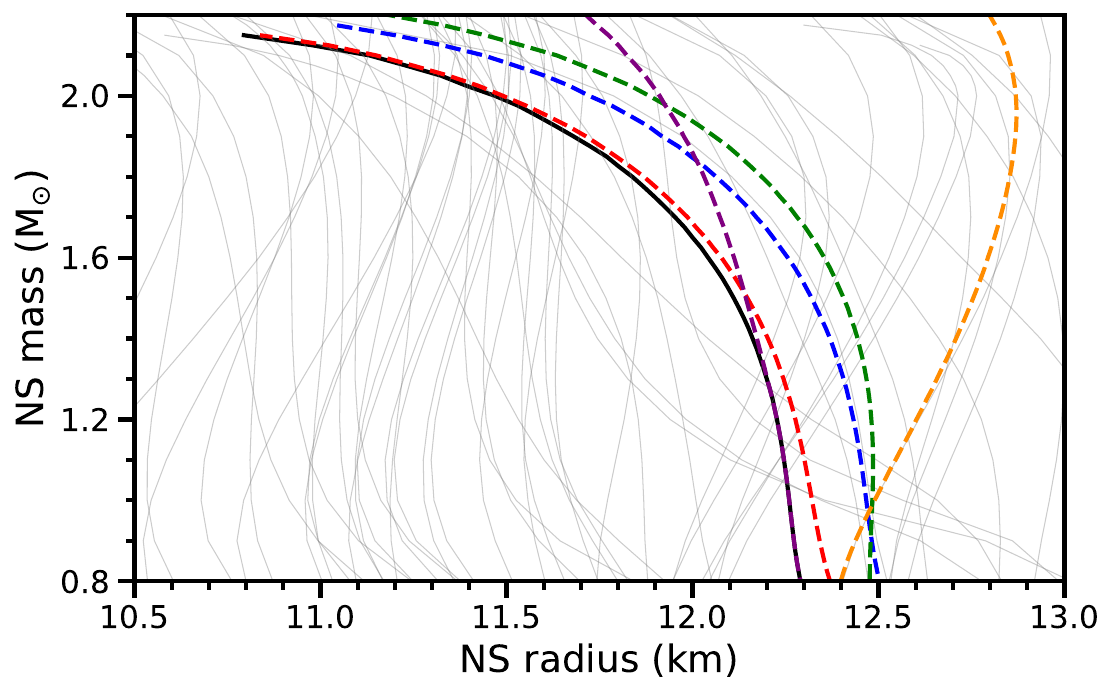}
\includegraphics[width=0.45\textwidth,angle=0]{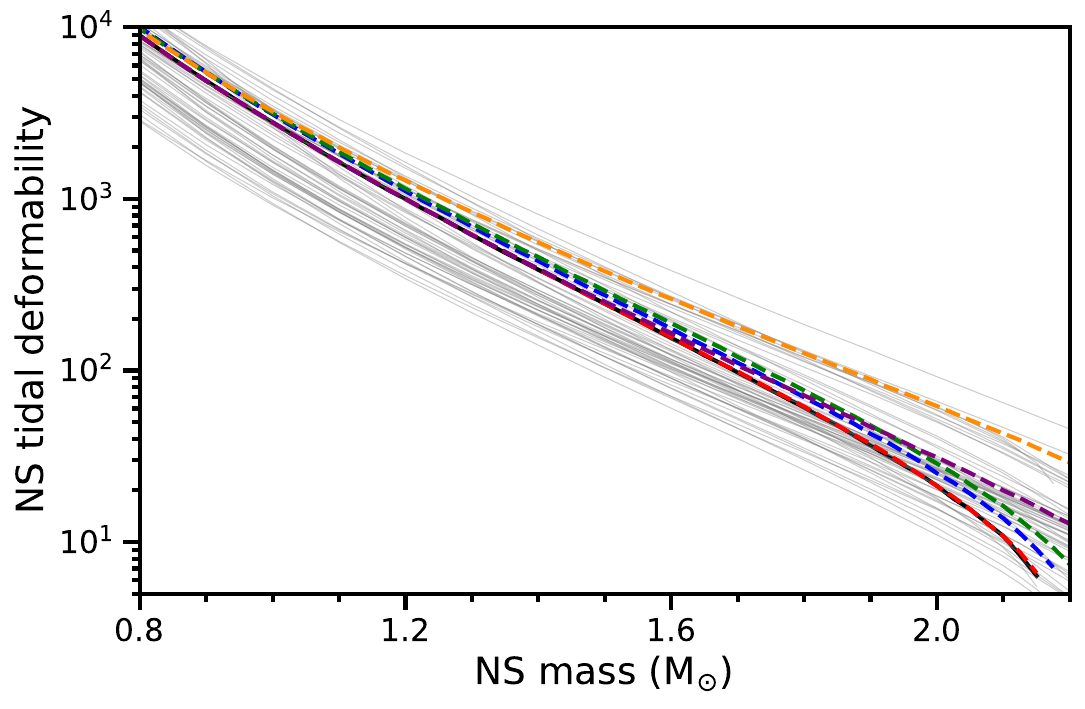}
\caption{The $i$-mode frequency, radius, and tidal deformability as functions of NS mass for various NS models, showing the dependences of these properties and their relationships with NS mass on the choice of model. The $i$-mode frequency's mass dependence appears to be similar for all of the models, in contrast to the radius's more complex dependence. Six models are highlighted, with the solid black lines being for the model generated with our injected meta-model parameter values ($J=31\text{ MeV}$, $L=50\text{ MeV}$, $K_{\rm sym}=-100\text{ MeV}$, $\gamma_1=3.0$, $\gamma_2=3.0$). The dashed lines give some indication of how significant the model parameters are for these NS properties by changing one of the parameters by a small amount while keeping the others at our injected values: blue has $J=32\text{ MeV}$, red has $L=60\text{ MeV}$, green has $K_{\rm sym}=-50\text{ MeV}$, orange has $\gamma_1=4.0$, and purple has $\gamma_2=4.0$.}
\label{fig:mfrt_vary1}
\end{figure}

In Figure~\ref{fig:mfrt_vary1} we plot the relationship between $i$-mode frequency and NS mass for our injected NS model, and the relationships obtained by individually changing each of the injected parameters by a small amount. To provide some context for how varying the meta-model parameters affects the resulting NS model, we also plot the relationships between NS mass and radius and between NS mass and tidal deformability, to show how changing the parameter values affects these more commonly considered NS properties. We can see that the mass-frequency relationship is near-linear and has a shape that only changes slightly for different sets of parameters.

The changes are small when compared to the sensitivity of the mass-radius relationship to those parameters, and from the wider range of relationships shown in grey it is clear that there is much more freedom in the shape of the mass-radius relationship than the mass-frequency relationship. For a given EOS, the mass-frequency relationship is nearly linear with the difference in $i$-mode frequency across the full range of NS masses being similar to changes caused by small variations of the meta-model parameters and less than the frequency uncertainty we may expect from coincident timing \citep{tsang2012resonant}. It is therefore unclear whether uncertainty in NS mass will have a significant effect on any parameter constraints inferred from measurements of the $i$-mode frequency, and whether combining multiple measurements to probe the mass-frequency relationship would be useful, and motivates our work in this study.

\section{Masses of NSs and BHs in merging binaries} \label{sec:population}
\subsection{Artificial set of binary mergers}
To investigate whether combining $i$-mode frequency measurements from several multimessenger RSF and GW events could allow us to constrain neutron star structure and the nuclear symmetry energy better than with a single measurement, we must first construct an artificial set of binary systems to serve as the progenitors of injected events. The rate at which we may observe multimessenger RSF and GW events is unclear. The upper limit estimated by~\citet{neill2022resonant} was $\sim 3$ events per year (combining those from BHNS and NSNS binaries). This suggests that while they may not be rare, we should not expect a very large number of these mergers, in agreement with the current absence of any such multimessenger detections. We therefore choose to have 5 binaries in our set of artificial systems: a small number that is not too unreasonable a hope for observations in the next decade or so. 

Assuming that there is a single NS EOS and that all NSs have time to spin down and cool to negligible temperatures prior to binary merger, any differences between the $i$-modes of NSs will be related to their masses, and so we shall focus on the masses of the NSs in our binary systems. To have our injected events be consistent with currently known mergers, we draw these masses from the extragalactic NS mass distribution inferred by \citet{Landry2021Mass} from observed GW signals produced by compact binary mergers involving NSs (specifically, their ``FLAT'' model). While the small number of such events means that this distribution is unlikely to accurately predict future detections, it will be sufficient for this work. Using a NS population inferred from GW observations will bias our NS masses towards larger values since they produce louder GW signals, but that is appropriate for this work as we are only interested in binary mergers that could be detected in GWs. We do not know of any correlation between RSF detectability and NS mass, as their emission mechanism has not been studied in detail.

The mass of the binary companion to a NS RSF progenitor will strongly affect their GW signal and thus influence the uncertainty in the NS's mass, so we must also select masses for the companions in each binary we use.
We will consider both BHNS and NSNS binaries, as -- unlike SGRBs or Kilonovae -- RSFs can be produced by NSs in BHNS binaries as easily as those in NSNS binaries~\citep{neill2022resonant}. The binary population synthesis of BPASS~\citep{eldridge2017binary,stanway2018reevaluating} finds that there is approximately one BHNS for every four NSNS systems, and while GW signals from more massive binary mergers are detectable at greater distances, the luminosities of RSFs may not have such a strong dependence on mass. As the distances at which RSFs may be visible~\citep{neill2022resonant} are similar to the current GW detection range for NSNS mergers~\citep{abbott2020prospects}, we therefore assume that RSFs produced by BHNS mergers beyond the NSNS range will not be detected. This means that approximately $20\%$ of multimessenger events will originate from BHNS binaries, and so we will have one of our five random binaries be a BHNS system. While we draw NS masses from \citet{Landry2021Mass}'s extragalactic NS mass distribution, for the BH's mass we randomly select a value from the range $3$-$30\text{ M}_{\odot}$. We ignore any correlations that may exist between the masses of the two objects in a binary.

Aside from combining $i$-mode frequency measurements for NSs that have a realistic distribution of masses, we shall also examine the extreme case where multimessenger RSF and GW events are detected for NSs with very high and low masses. Combining frequency measurements from these extremes will provide the strongest discriminating power for small differences between the slopes of the (near-linear, see Figure~\ref{fig:mfrt_vary1}) mass-frequency relationships of different NS models. This case will therefore maximise the benefit of a obtaining second measurement for probing the mass-frequency relationship. The NS with the lowest known mass is the binary partner of the pulsar J0453+1559, which has a mass of $1.174^{+0.004}_{-0.004}\text{ M}_{\odot}$ \citep{Martinez2015Pulsar}. Lower mass NSs would be physically stable~\citep{Haensel2002Equation}, but it is doubtful that a much lower mass NS could be produced, as most formation channels will favour producing white dwarfs for masses well below the Chandrasekhar limit. The maximum NS mass is a topic of much interest as it is important for the NS population and is closely related to the high density behaviour of NS EOS. The most massive NS with a well-determined mass is the pulsar J0740+6620 with mass $2.08^{+0.07}_{-0.07}\text{ M}_{\odot}$ \citep{Fonseca2021Refined}, and other astrophysical constraints typically place the limit between $2.0\text{ and }2.3\text{ M}_{\odot}$ \citep[see, e.g.][]{margalit2017constraining,rezzolla2018using}. For our extreme NS masses we choose to use binaries with $m_2=1.0\text{ M}_{\odot}$ and $m_2=2.0\text{ M}_{\odot}$ NSs, where $m_2$ is the mass of the lighter object in the binary. We shall not investigate how the masses of their companions affect the results and simply set them to $m_1=m_2+0.1\text{ M}_{\odot}$.

It may be possible for both of the NSs in a NSNS binary to produce a RSF, resulting in two RSFs separated by a duration determined by how the difference in their masses affects their $i$-mode frequencies. However, mechanisms by which a bright flare could be produced when the NS crust resonantly shatters typically require the NS to have a strong \textit{surface} magnetic field \citep{tsang2012resonant,neill2022resonant}. Most NSNS systems will have been born and evolved as a binary, and for a binary to merge within the Hubble time it must be fairly tight, meaning that the first star to collapse and become a NS will likely accrete a significant amount of matter during the evolution of its binary partner. NSs that accrete a significant amount of matter will have their magnetic fields buried \citep{Alpar1982, Bhattacharya1991, Cumming2001}, making it unlikely that they will be capable of producing RSFs. In the majority of NSNS binaries the NS that formed first will be the more massive one, and so we shall assume that the less massive NSs in our binaries are always the ones to produce the RSFs, while the more massive ones do not. Two RSFs could be produced by a binary that formed dynamically in a dense stellar environment \citep{Ye2020Globular}, or from a binary that initially had a large separation such that no accretion took place but was later 'kicked' \citep{ghodla2021forward} into a closer orbit, but these scenarios will be rare and so we shall not consider them here.

\subsection{NS mass uncertainty from GW inference}\label{sec:bilby}
For a real multimessenger event the masses of the objects in the binary will not be well known, as while GWs allow for a precise measurement of the binary's chirp mass,
\begin{equation}
\mathcal{M}=\frac{m_1^{3/5}m_2^{3/5}}{m_1^{1/5}+m_2^{1/5}},
\label{eq:chirp_mass}    
\end{equation}
\noindent they do not strongly constrain the masses of the individual objects. However, the dependence of $i$-mode frequency on NS mass shown in Figure~\ref{fig:mfrt_vary1} is strong enough that the frequency range corresponding to a large mass uncertainty could be comparable to the uncertainty in $i$-mode frequency values measured using coincident GW and RSF timing \citep{tsang2012resonant}. When inferring the parameters of our NS meta-model using such frequency measurements we should therefore account for the uncertainty in the mass of the NS which produced the RSF.

We could try to estimate reasonable mass uncertainties by examining the results of real GW analyses, but currently only a small number of binary mergers involving NSs have been detected~\citep{abbott2019properties,abbott2020gw190425,abbott2021observation} and so such estimates might not be representative of all mergers. Instead, to get estimates of the uncertainties in the masses of objects involved in binary mergers, we use \texttt{Bilby}~\citep{Ashton2019BILBY} to obtain realistic posterior probability distributions for the analysis of those binary's GW signals. \texttt{Bilby} is a Python module for performing Bayesian inference, with its main focus being GW binary parameter inference. It also contains functionality to generate synthetic GW signals from a provided set of binary parameters, allowing us to inject our artificial binaries and then investigate the precision to which their masses can be recovered.

We assume that preliminary analysis of a GW signal will allow for a strong prior on its chirp mass and consequently determine whether it originates from a NSNS binary with reasonable confidence. To generate the synthetic GW signal for a binary and perform parameter recovery we must inject a set of extrinsic and intrinsic parameters (i.e. parameters that are and are not dependent on the observer's position relative to the binary system). The full parameter space of this is very large, and so rather than recovering every parameter we reduce computation times by fixing some parameters that we do not expect to have significant degeneracy with NS mass to their injected values. The parameter which we do not fix are listed in Table~\ref{tab:GW_params}, alongside the priors we use for their recovery. As is typical for GWs, rather than directly inferring the separate masses of the objects in the binary we infer the mass ratio and chirp mass, as they more directly affect GW signals. When generating synthetic GW signals, we assume detections involving the LIGO (Livingston and Hanford) and Virgo interferometers, and use estimates of their O4 sensitivities \citep{abbott2020prospects}\footnote{\url{https://dcc.ligo.org/LIGO-T2000012/public}}. Note that, in light of current detector status, these sensitivities might only be achieved in the second half of O4 or during O5.

Aside from the mass ratio and chirp mass, the other intrinsic parameters we recover are spin and tidal deformability. The spin of the more massive object in the binary is included in our inferences with the same low-spin prior as \citet{abbott2017gw170817}, which is based on not having observations of NS in close binaries with high spins. We do not include the spin of the less massive object, because in order to produce a bright RSF it must have a strong magnetic field~\citep{neill2022resonant} and such a field will cause its spin to decay to near-zero long before merger. It is worth noting that in GWs there can be significant degeneracy between mass and spin, and so broader spin priors could increase mass uncertainty \citep[e.g.][]{abbott2017gw170817}, but we do not concern ourselves with that here. For simplicity we also assume that the spin and orbit axes are perfectly aligned.

Tidal deformability is the property of NS matter that has the lowest-order effect on GW emission. For each of our NSs we use our injected EOS to calculate its injected value, which scales with NS mass as $m^{-5}$. This strong mass-dependence means that the reasonable range for tidal deformability values varies by up to an order of magnitude for different NS masses, and so using the same prior at all masses would result in it being either too tightly constrained at low mass or a too broad at high mass, making the posteriors sensitive to the choice of prior. We therefore choose to use a mass-dependent tidal deformability prior. Using a large selection of samples from our NS meta-model over wide parameter ranges, we find that the range of mass-tidal deformability relationships is contained within the bounds
\begin{align}
500m^{-4-m}\leq\Lambda(m)\leq10000m^{-4-m}
\label{eq:td_uniform_prior}
\end{align}
\noindent where $m$ is the NS mass in units of solar masses. The distribution of mass-tidal deformability relationships from these samples is non-uniform and might be better fit by a truncated normal distribution, but for simplicity we shall just use a uniform prior between these bounds. We use equation~\eqref{eq:td_uniform_prior} for both $\Lambda_1$ and $\Lambda_2$, ignoring that their values should be correlated for two NSs that share the same EOS. $\Lambda_1$ is of course fixed to zero for BHs.

\begin{table*}
\centering
\addtolength{\tabcolsep}{-0.3pt}
\renewcommand{\arraystretch}{1.1}
\begin{tabular}{| c | c | c |}

\hline

Intrinsic parameter & Symbol & Prior distribution  \\
\hline
Mass ratio & $q$ & Uniform from $0.125$ to $1.0$ \\
Chirp mass & $\mathcal{M}$ & Uniform in component masses from $\mathcal{M}_{\rm in}-0.2\text{ M}_{\odot}$ to $\mathcal{M}_{\rm in}+0.2\text{ M}_{\odot}$  \\	
Tidal deformability of 1st object & $\Lambda_1$ & Uniform from $500m_1^{-4-m_1}$ to $10000m_1^{-4-m_1}$ \\
Tidal deformability of 2nd object & $\Lambda_2$ & Uniform from $500m_2^{-4-m_2}$ to $10000m_2^{-4-m_2}$ \\
Higher mass object's aligned spin & $\chi_1$ & Uniform from $0$ to $0.05$ \\

\hline

Extrinsic parameter & Symbol & Prior distribution  \\
\hline
Polarisation angle & $\psi$ & Uniform from $0$ to $\pi$ \\
Orbital phase at coalescence & $\phi$ & Uniform from $0$ to $2\pi$ \\
Time of coalescence & $t_c$ & Uniform from $t_{c,\rm in}-0.1$ to $t_{c,\rm in}+0.1$ \\
\hline
\end{tabular}

\caption{The parameters included in our Bayesian analyses of NSNS merger GW signals with \texttt{Bilby}, and their prior probability distributions. The subscript ``$\rm in$'' indicates the injected value of the parameter. The prior for $\Lambda_1$ is only for NSNS binaries, as its value is fixed to zero for BHNS binaries. In addition to these parameters, the prior is shaped by separate constraints on the masses of each object in the binary (see text).}
\label{tab:GW_params}
\end{table*}

The injected tidal deformability values are obtained from our injected EOS, but for the other parameters listed in Table~\ref{tab:GW_params} we randomly draw values to inject from their priors.
Aside from the parameters we sample, other properties such as the binary inclination, sky location and luminosity distance need injected values, which we draw from uninformed probability distributions: the isotropic $p(i)=\sin(i)$ for binary inclination, uniform-in-sky for sky location, uniform-in-volume for luminosity distance, and so on. However, the extrinsic parameters -- in particular luminosity distance and binary inclination -- strongly affect the GW signal-to-noise ratio (SNR). Therefore, as we are only interested in mergers that are detected in GWs and not those that are too weak to be clearly detected, for each injected event we repeatedly sample sets of extrinsic parameters until we find a set for which the SNR is $\gtrsim 8$ in all three LIGO and Virgo interferometers, comfortably above the threshold for a GW trigger~\citep{Abbott2021GWTC-3}. We also restrict the luminosity distance to below $200\text{ Mpc}$, as beyond this distance NSNS mergers will rarely have high SNRs.

\begin{figure*}
\centering
\includegraphics[width=0.9\textwidth,angle=0]{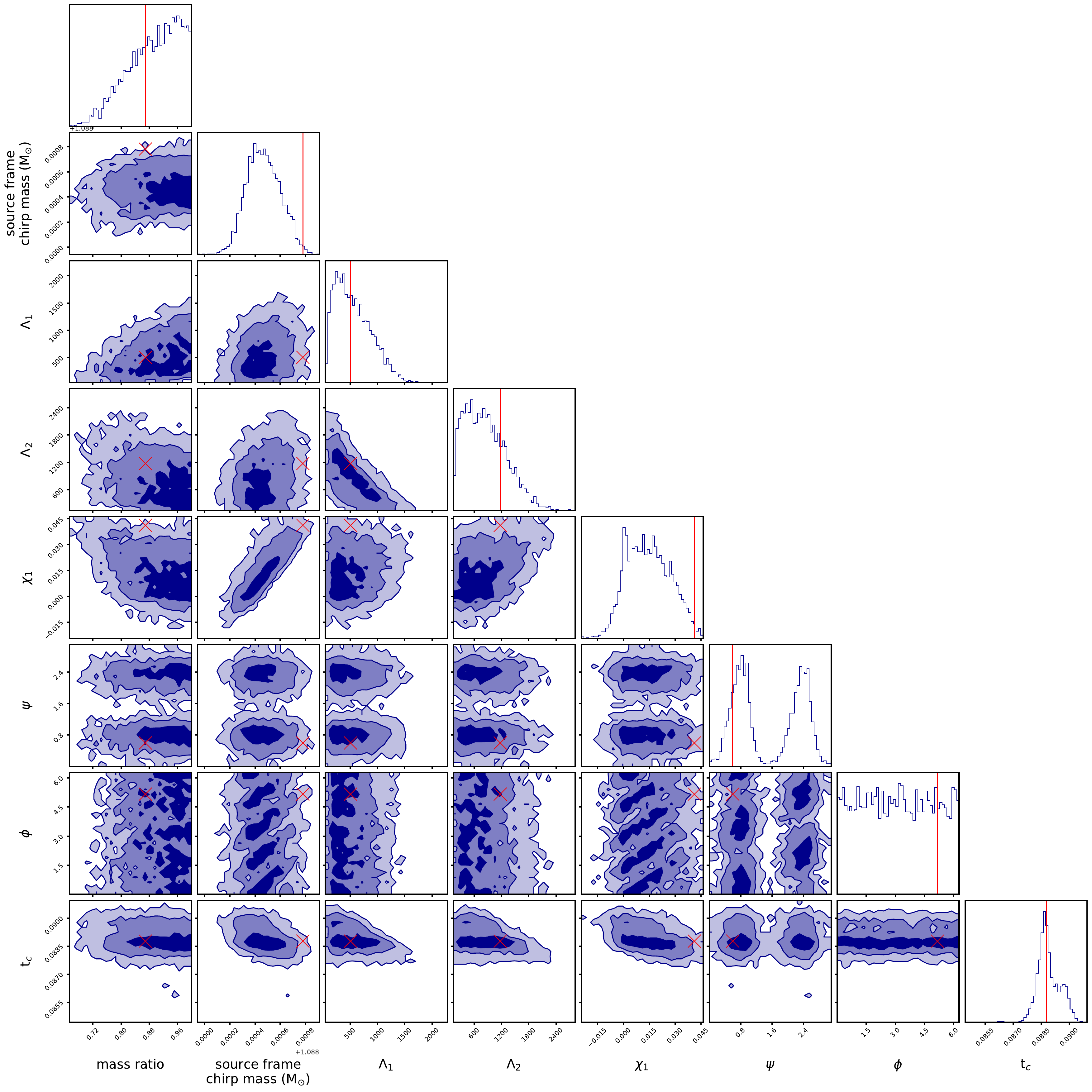}
\caption{Example posteriors for a NSNS binary's parameters recovered using Bayesian inference of an injected GW signal with \texttt{Bilby}, showing the scale of the uncertainties we can expect from such a signal. The first four parameters describe the masses and tidal deformabilities of both objects in the binary, and the fifth is the spin of the more massive one. The remaining three are extrinsic (observer-dependent) parameters. Other parameters could be relevant to GW inference, but for computational expediency they have been fixed to their injected values during this inference. The injected values of the inferred parameters are shown as red lines or markers on each plot. The priors used are given in Table~\ref{tab:GW_params}.}
\label{fig:10paramBilby1}
\end{figure*}

We use the \texttt{IMRPhenomPv2\_NRTidal} and \texttt{IMRPhenomNSBH} waveform models to inject and recover NSNS and BHNS mergers, respectively. For computational expediency we use \texttt{Bilby}'s implementation of the relative binning method~\citep{Krishna2023Accelerated} with the precision parameter $\epsilon=0.001\text{ radians}$, which allows for the inferences to be completed in hours instead of days or weeks.
Posteriors for one of our artificial binaries, which consists of NSs with masses $1.167$ and $1.342\text{ M}_{\odot}$, are shown in Figure~\ref{fig:10paramBilby1}. Comparing these to the posteriors of an inference for the same GW signal without relative binning, the difference is small, so this precision seems reasonable. As expected, the chirp mass of the binary is well constrained while the mass ratio (and thus the separate masses of the NSs) is more uncertain. Some uncertainty is introduced to the chirp mass through its correlation with the poorly recovered spin $\chi_1$, but it is still very accurately determined. However, if a prior were used which allowed higher spins -- which would be more reasonable in the case of, for example, a dynamically formed NSNS binary -- then this correlation could be more impactful. Our other injected binaries result in posteriors that are qualitatively similar to those shown in Figure~\ref{fig:10paramBilby1}.

\begin{figure}
\centering
\includegraphics[width=0.45\textwidth,angle=0]{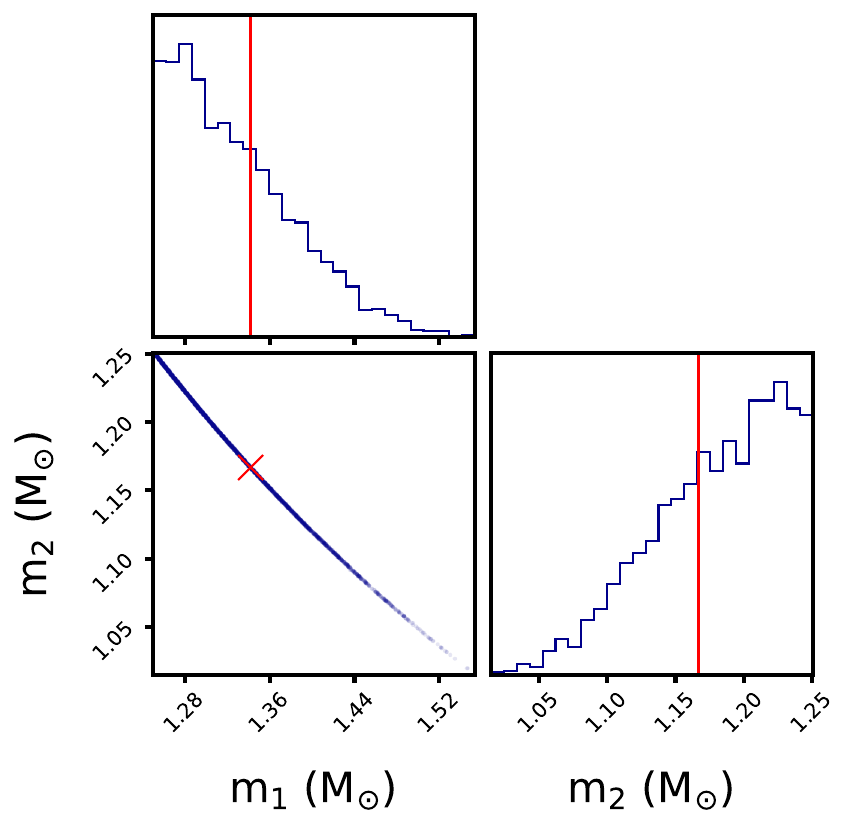}
\caption{The NS mass posteriors corresponding to the chirp mass and mass ratio posteriors from Figure~\ref{fig:10paramBilby1}. This $m_2$ posterior is the mass uncertainty we use for this NS in our meta-model parameter inferences in Section~\ref{sec:RSF_inference}.}
\label{fig:Bilby1_masses}
\end{figure}

We convert the mass ratio and chirp mass posterior samples used in Figure~\ref{fig:10paramBilby1} into the masses of the individual NSs, which are shown in Figure~\ref{fig:Bilby1_masses}. 
These are reasonably well constrained, having uncertainties of $\sim0.1\text{ M}_{\odot}$, and the inferences for our other injected binaries give similar uncertainties. From Figure~\ref{fig:mfrt_vary1} we can see that over a range of a few times $0.1\text{ M}_{\odot}$ the change in $i$-mode frequency is not large relative to changes from small variations of our meta-model's parameters, which could indicate that mass uncertainty will not affect the results of parameter inference using $i$-mode frequency measurements. In the next section we shall perform such inferences to see if this is the case.

\section{Parameter inference using mass-dependent \textit{i}-mode frequency measurements}\label{sec:RSF_inference}

\subsection{Prior parameter distribution and current constraints}\label{sec:prior_and_nuclear}

We begin constructing our prior by selecting broad parameter ranges: $25<J<43\text{ MeV}$, $5<L<158\text{ MeV}$, $-520<K_{\rm sym}<200\text{ MeV}$, $1.001<\gamma_1<20\text{ MeV}$ and $1.001<\gamma_2<20\text{ MeV}$. We reject any section of this parameter space which produces an unstable nuclear matter equation of state. To this we add the constraint that the NS maximum mass must satisfy $2.1<M_{\rm max}<2.5\text{ M}_{\odot}$. The lower bound of $2.1\text{ M}_{\odot}$ is above the mass ranges we obtain from GW inference for all of our artificial binaries, meaning that our prior already contains all of the maximum NS mass information these GW analyses would provide. Our inferences using $i$-mode frequency data in this Section will therefore not be contaminated by also containing new information about the NS maximum mass, isolating the benefit of probing the $i$-mode frequency. The upper bound sits comfortably above the values suggested by current observations \citep{Fonseca2021Refined,Romani2022J0952} and helps to limit the parameter space of our meta-model that produces more improbable models. These maximum mass constraints mainly affect the ranges of our meta-model's core polytrope parameters, as the maximum mass is sensitive to the high-density equation of state.

The polytropes that we use in the core result in the speed of sound ($c_s$) continually increasing with density, and have no physical content to prevent it from becoming greater than the speed of light. To ensure that this most basic of physical requirements is satisfied, if the parameter values of a sample from the prior would result in the speed of sound becoming greater than the speed of light at a density that can be found in NSs -- i.e. anywhere below the central density of the maximum mass NS -- then we simply set its prior probability to zero. This approximately translates to rejecting samples with $\gamma_1\gtrsim5$, and with high $\gamma_2$ when $\gamma_1$ is not very low.

The prior we have constructed is very wide, and our knowledge of the symmetry energy is better than that as multimessenger RSF and GW events are of course not the only way to constrain the symmetry energy parameters. We shall therefore also consider various experimental constraints on NSs and nuclear matter, to which we can then add information from RSFs. We list the nuclear physics constraints we consider in Table~\ref{tab:nuclear}. These constraints are from combining the results of various experimental in quadrature \citep{newton2021nuclear}, and so the errors given in the Table's caption are optimistic lower bounds. We use each of the listed values as the mean of a normal distribution, with its error as its standard deviation. Our nuclear physics likelihood for a sample from the prior is then the product of all of the normal distributions at the values produced by that sample.

\begin{table}
    \centering
	\addtolength{\tabcolsep}{-0.3pt}
    \renewcommand{\arraystretch}{1.1}
    \begin{tabular}{| c | c | c | c | c |}
        
        \hline
        
		Nucleus & $E_B$ (MeV) & $r_{ch}$ (fm) & $r_{np}$ (fm) & $\alpha_D$ (fm$^3$)  \\
		\hline
		$^{16}\text{O}$   &  127.62 & 2.699 &       &      \\
		$^{40}\text{Ca}$  &  342.05 & 3.478 &       &      \\
		$^{48}\text{Ca}$  &  415.99 & 3.477 & 0.195 & 2.07 \\
		$^{56}\text{Ni}$  &  483.99 & 3.477 &       &      \\
		$^{132}\text{Sn}$ & 1102.85 & 4.709 &       &      \\
		$^{208}\text{Pb}$ & 1636.43 & 5.501 & 0.178 & 19.6 \\
        \hline
	\end{tabular}
	\caption{Nuclear data we include in our inferences to incorporate our current understanding of the symmetry energy parameters. From left to right: binding energies, charge radii, neutron skin thicknesses, and dipole polarisabilities. We take the errors on the binding energies and charge radii to be $3\text{ MeV}$ for all binding energies, $0.02\text{fm}$ for all charge radii. We $0.013$ and $0.011\text{ fm}$ ($0.22$ and $0.6\text{ fm}^3$) for the neutron skins (dipole polarisabilities) of $^{48}\text{Ca}$ and $^{208}\text{Pb}$, with the neutron skin errors from a number of measurements compiled in \citet[and references therein]{newton2021nuclear}.}
    \label{tab:nuclear}
\end{table}

The main effect of this nuclear physics data is to constrain the lower order symmetry energy parameters, and in particular $J$. This can be seen from the results plotted in grey in Figure~\ref{fig:Muncertainty}, where the $J$ range is substantially smaller than the $25$ to $43$ MeV range chosen for our prior. We can also see that correlations appear between the $J$ and $L$ parameters, and between the $L$ and $K_{\rm sym}$ parameters. Nuclear masses give their most stringent constraints a little above half saturation density. If $J$ is higher, then in order to reach the same value of the sub-saturation symmetry energy, the slope of the symmetry energy must be steeper - i.e. $L$ must be larger. Likewise, for a given value of $J$, if we increase $L$, then the slope must change more rapidly to maintain the same value of the sub-saturation symmetry energy. This results in the positive correlations seen in Figure~\ref{fig:Muncertainty} \citep{lattimer2013constraining}.

\subsection{Injected \textit{i}-mode frequency measurements}\label{sec:injected_f}
Now that we have results for current constraints, we can add measurements of NS properties from multimessenger GW and RSF events to examine how they might improve our understanding of NSs and nuclear matter. Multimessenger RSF and GW events allow for measurements of the $i$-mode frequency within their progenitor NSs, which can be used to infer the structure and composition of the NS crust~\citep{Neill2023Constraining}. A RSF occurs when a NS's (quadrupolar) $i$-mode is resonant with the binary orbit, and the frequency of (quadrupolar) GWs is twice the orbital frequency, meaning that the GW frequency at the time at which a RSF occurs ($f_{\rm RSF}$) will be approximately equal to the frequency of that NS's $i$-mode ($f_{2i}$). However, the finite duration of a RSF means that a measurement of the $i$-mode frequency from coincident GW and RSF timing will include some uncertainty, on the order of the change in GW frequency over the duration of the flare.
For each injected binary system we therefore estimate this uncertainty to construct a probability distribution for the $i$-mode frequency.

We choose to use a normal distribution for each frequency measurement, centred on the value calculated using our injected NS model for a NS with mass $m_2$ (i.e., that NS's ``true'' $i$-mode frequency). We calculate $i$-mode frequency values by combining the outputs of the TOV equations with \citet{yoshida2002nonradial}'s relativistic pulsation equations.
For the $i$-mode frequency distribution's standard deviation, we use the change in GW frequency during the $i$-mode's resonance~\citep{tsang2012resonant}:
\begin{equation}
\delta f\sim t_{\rm res}\frac{\partial f_{\rm GW}}{\partial t}\bigg\rvert_{f_{\rm RSF}}\mkern-18mu \sim11.19\text{Hz}\left(\frac{\mathcal{M}}{1.2\text{ M}_{\odot}}\right)^{\frac{5}{6}}\left(\frac{f_{\rm RSF}}{200 \text{ Hz}}\right)^{\frac{11}{6}}.
\label{eq:freq_spread}    
\end{equation}
\noindent The likelihood that a NS with mass $m_{\rm NS}$ following the EOS corresponding to the parameter values $\theta\equiv\left\{J,L,K_{\rm sym},\gamma_1,\gamma_2\right\}$ produced an injected RSF is therefore
\begin{equation}
L\left(f_{\rm RSF}|\theta,m_{\rm NS}\right)=\exp\left(-\frac{1}{2}\frac{\left(f_{2i}\left(\theta,m_{\rm NS}\right)-f_{\rm RSF}\right)^2}{\delta f^2}\right)
\label{eq:Likelihood_normal}
\end{equation}
\noindent where for our injections $f_{\rm RSF}=f_{2i}(\Phi,m_2)$, with $\Phi$ being the injected parameter values.

\begin{figure}
\centering
\includegraphics[width=0.45\textwidth,angle=0]{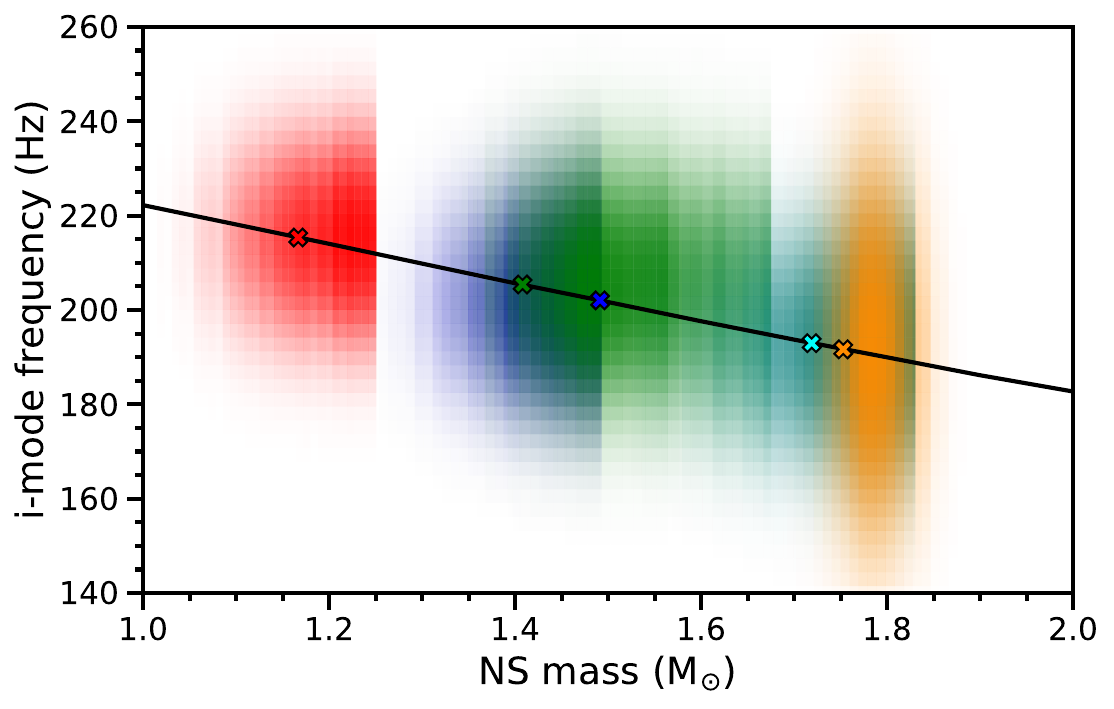}
\caption{The injected data we use for the masses and $i$-mode frequencies of five NSs, showing the sort of information we can expect to obtain from multimessenger RSF and GW events. The evidence from each merger is shown in a different colour, with darker shades indicating greater evidence. The injected $i$-mode frequency and mass values are shown as markers (with the same colours as their evidences), which lie along the injected NS model's mass-frequency relationship.}
\label{fig:fm_evidence}
\end{figure}

In Figure~\ref{fig:fm_evidence} we combine the uncertainty in $i$-mode frequency from coincident timing with the uncertainty in $m_2$ from GW analysis to show how our 5 injected mergers with random masses inform us of the injected mass-frequency relationship. We can see that the BHNS binary (orange) gives qualitatively different mass and frequency information to the NSNS binaries (red, green, blue, cyan). The BHNS has a larger frequency uncertainty because its higher chirp mass causes a higher rate of change of orbital frequency during resonance, while the NSNS binaries' mass distributions have sharp cutoffs at half their total masses. This is because total mass is well constrained by GW analysis, meaning that there is a strong upper limit on the mass of the less massive object in a binary. The NSNS binaries are all close enough to being equal mass that this affects their posteriors, while the BHNS binary is clearly not near equal mass. However, the changes in the mass-frequency relationship across the mass uncertainty ranges are small for all of these binaries. Given that this relationship is similar for all of the NS models shown in Figure~\ref{fig:mfrt_vary1}, we therefore expect that the relative likelihoods of different NS models will be similar to the relative likelihoods obtained when fixing the masses to their injected values.

Each $i$-mode frequency measurement included in an inference that is taken from a different RSF introduces additional parameters: the parameters of the binary that can affect the NS $i$-mode frequency. In this work we assume that the NS's mass is the only one of these parameters that could significantly affect the $i$-mode, as while properties such as the NS's spin and temperature can affect the $i$-mode we expect that they will be negligible (approximately zero in the cases of spin and temperature) for any NS that produces a RSF. However, adding even one extra parameter per $i$-mode frequency measurement means that the computation time required to sample the parameter space will increase significantly as more measurements are considered. To keep our computation times reasonable for this work, we avoid increasing the parameter space by not fully including the NS masses as parameters to be inferred. Instead, for each measurement we marginalise over the mass parameter in Bayes equation to find the likelihood function:
\begin{align} 
L\left(f_{\rm RSF,i}|\theta\right)=\int L\left(f_{\rm RSF,i}|\theta,m_{2,i}\right)\pi\left(\theta,m_{2,i}\right)dm_{2,i},
\label{eq:manyRSF_marginalise}
\end{align}
\noindent where the integral is over the range of \texttt{Bilby}'s posterior for $m_2$ in this binary ($p_{\rm GW}(m_2)$), and $\pi$ is the probability distribution given by
\begin{align}
\pi\left(\theta,m_{2,i}\right)=\begin{cases}
p_{\rm GW}(m_2) \;\;\text{ if } m_2\leq m_{\rm max}(\theta)\\
\:\;\;\;\;\; 0 \;\;\;\;\;\;\;\;\;\text{ if } m_2> m_{\rm max}(\theta).
\end{cases}
\end{align}
\noindent This marginalisation over mass means that the posteriors from our inferences will not show any correlations that might exist between the masses and the other parameters, but the weakness of the mass-frequency relationship means that such correlations are unlikely to be significant.

When combining measurements of the $i$-mode frequency from several multimessenger RSF and GW events, the total likelihood function for our Bayesian inference is simply the product of likelihood functions for each event,
\begin{align} 
L\left(f_{\rm RSF,1},\ldots,f_{\rm RSF,N}|\theta\right)=\prod_{\rm i=1}^N L\left(f_{\rm RSF,i}|\theta\right),
\label{eq:manyRSF_likelihood}
\end{align}
\noindent since we can safely assume that binary mergers are isolated from each other. This total likelihood can then be used alongside the prior probability distribution 
\begin{align}
\pi(\theta)=\frac{1}{A_{\gamma_1\gamma_2}(J,L,K_{\rm sym})}
\end{align} 
\noindent which we introduced in Section~\ref{sec:prior_and_nuclear} to obtain the posterior distribution. To do this, we use the Markov chain Monte Carlo (MCMC) method via the \texttt{emcee} python module~\citep{foreman-mackey2013emcee}.

\subsection{Recovering the injected NS meta-model parameter values}

\begin{figure}
\centering
\includegraphics[width=0.45\textwidth,angle=0]{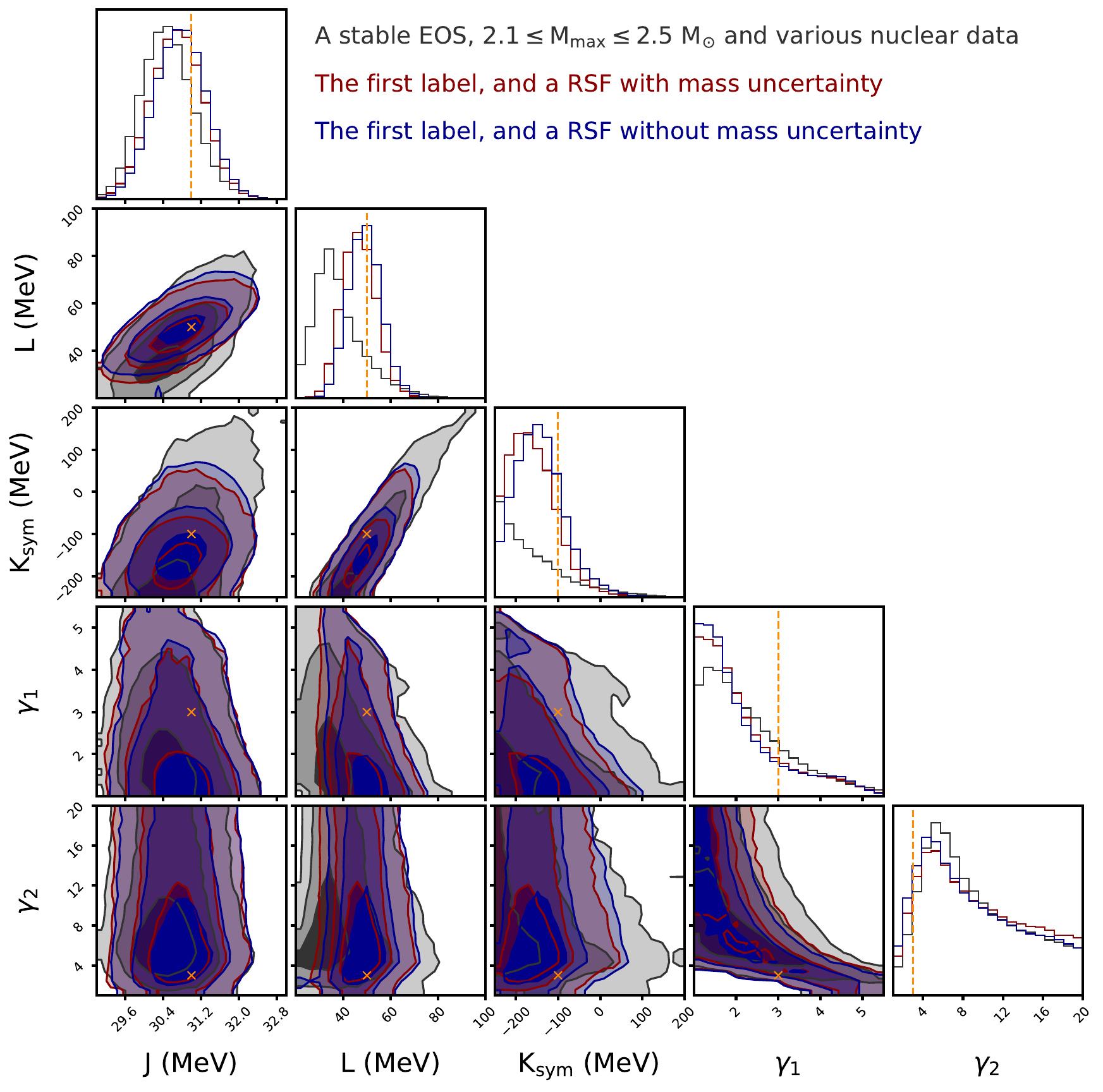}
\caption{Posteriors probability distributions from Bayesian inferences of our NS meta-model's parameters using different data, showing that $i$-mode frequency measurements from multimessenger RSF and GW events can improve on symmetry energy constraints from nuclear physics, and that including realistic uncertainty in NS mass has only a small impact on the results. The grey posteriors are for an inference using data from nuclear experiment and a constraint on the maximum NS mass, while the red and blue posteriors are for inferences that also included a single $i$-mode frequency measurement (the green one in Figure~\ref{fig:fm_evidence}) with and without mass uncertainty (respectively). The injected values are indicated by the orange markers and dashed lines.}
\label{fig:Muncertainty}
\end{figure}

We first examine whether the mass uncertainty from GW analysis can affect the results of meta-model parameter inferences that use $i$-mode frequency measurements. The posteriors of an inference using no frequency measurements (i.e., only including the experimental nuclear data) and of inferences using the measurement shown in green in Figure~\ref{fig:fm_evidence} with and without its mass uncertainty are shown in Figure~\ref{fig:Muncertainty}. From this we can see that mass uncertainty has little effect of the posteriors, even in the relatively extreme case of this particular data where the injected mass value is far from the center of the mass range inferred with \texttt{Bilby}. This Figure confirms that marginalising over mass in our likelihood function does not lose much information, as two different mass distributions giving almost identical posteriors indicates that correlations between the model parameters and the NS mass are insignificant.

\begin{figure}
\centering
\includegraphics[width=0.45\textwidth,angle=0]{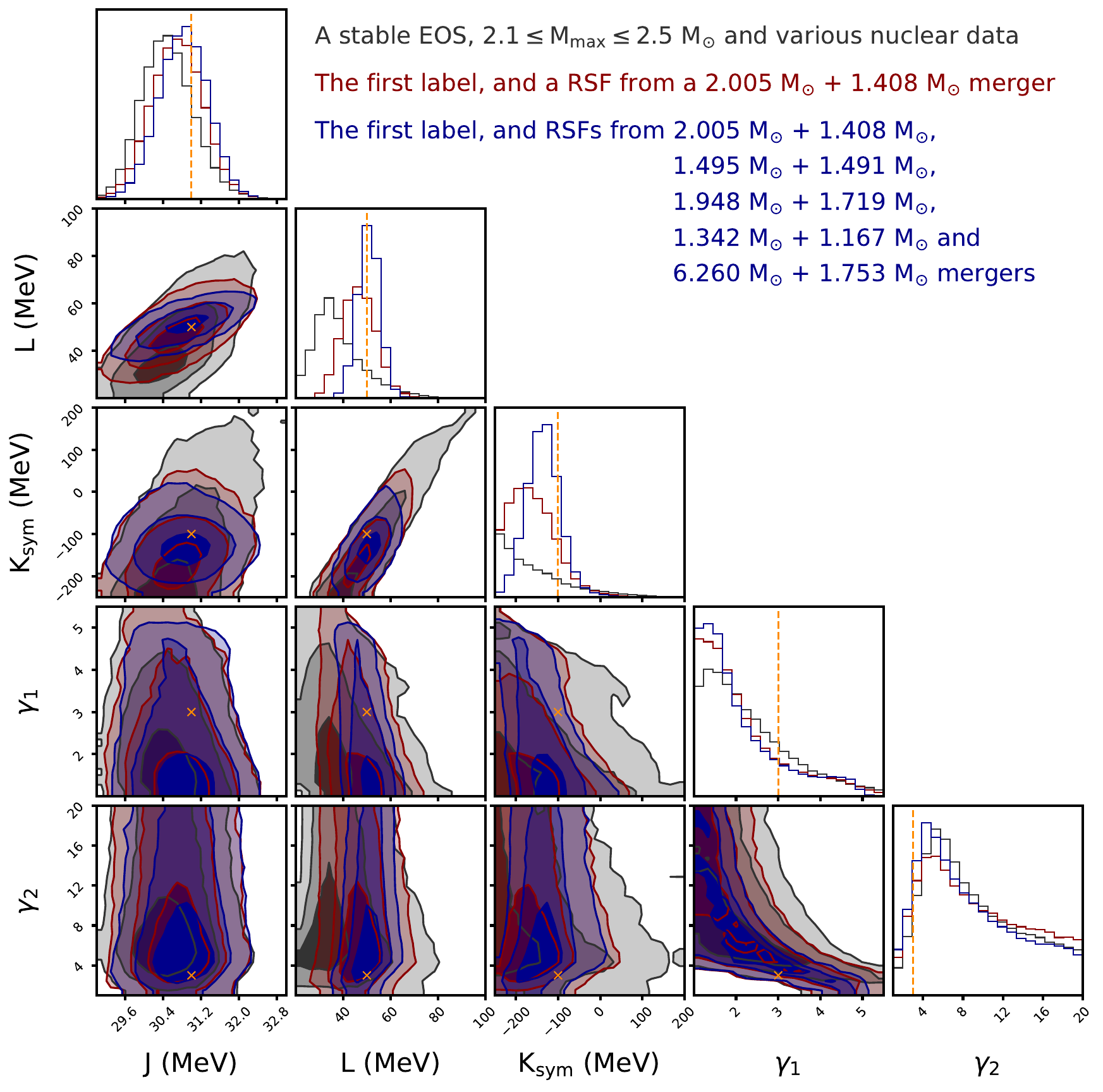}
\caption{Similar to Figure~\ref{fig:Muncertainty}, but for the posteriors of inferences using 1 (red) and 5 (blue) of the $i$-mode frequency measurements from Figure~\ref{fig:fm_evidence}, showing that measurements after the first continue to improve the symmetry energy constraints. Both inferences included NS mass uncertainty. The labels give the injected masses of the compact binary mergers for which the measurements were constructed.}
\label{fig:combineRSFs_4data}
\end{figure}

As found by \citep{Neill2023Constraining}, Figure~\ref{fig:Muncertainty} shows that a single $i$-mode frequency measurement can improve symmetry energy constraints obtained from nuclear experiment.
We next examine how combining measurements from several multimessenger events could further strengthen these constraints: posteriors for inferences using 1 and 5 injected frequency measurements (with mass uncertainty) are shown in Figure~\ref{fig:combineRSFs_4data}. From this we see that little information is gained about the core as the polytrope parameters change little with additional $i$-mode frequency data, but the symmetry energy is better constrained with additional measurements. While the posteriors for $J$ are similar as it is already well constrained by nuclear experiment, the $L$ and $K_{\rm sym}$ posteriors are substantially improved.

\begin{figure}
\centering
\includegraphics[width=0.45\textwidth,angle=0]{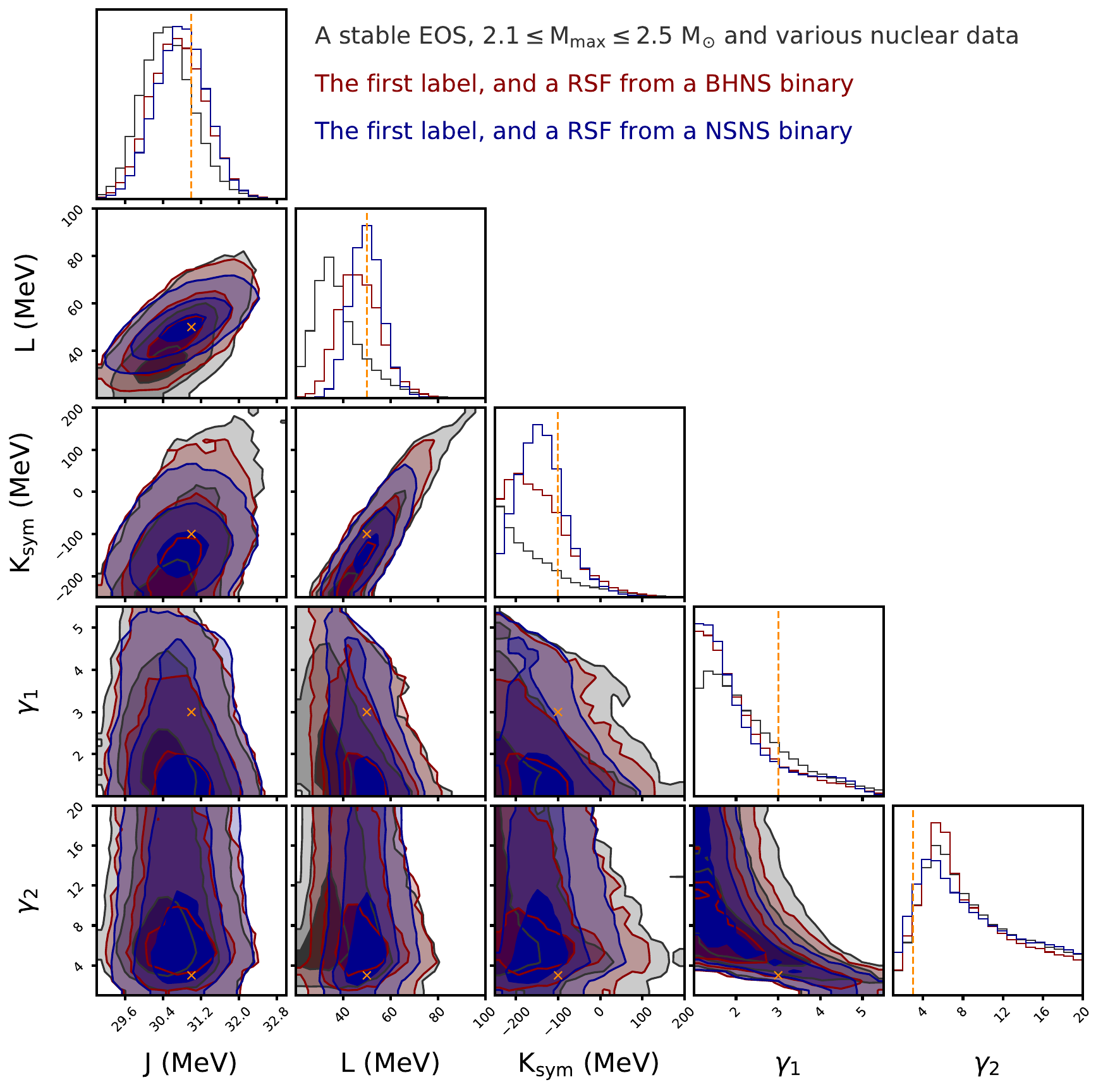}
\caption{Similar to Figure~\ref{fig:Muncertainty}, but comparing posteriors of inferences using $i$-mode frequency and NS mass measurements from a single BHNS merger (red) and from a single NSNS merger (blue). The NSNS merger appears to give better constraints, indicating that frequency uncertainty from coincident RSF and GW timing is more significant than how the $i$-mode frequency may change over realistic mass uncertainties. All properties of the injected binaries are identical except for the primary mass.}
\label{fig:combineRSFs_BHNS-NSNS}
\end{figure}

Figure~\ref{fig:combineRSFs_4data} includes data from four NSNS binaries and one BHNS binary, but in Figure~\ref{fig:fm_evidence} we saw that these different types of binary can give qualitatively different $i$-mode frequency measurements.
Therefore, to investigate whether symmetry energy constraints obtained from a RSF depend the type of binary which produced it, we compare the results of inferences using $i$-mode frequency measurements from a single NSNS binary and from a single BHNS binary. To eliminate other variables we inject two new binaries with identical extrinsic parameters and the same NS mass of $m_2=1.4 \text{M}_{\odot}$, but that have different masses for the binary partner: $m_1=1.6 \text{M}_{\odot}$ or $m_1=7.0 \text{M}_{\odot}$. The posteriors of their meta-model parameter inferences are shown in Figure~\ref{fig:combineRSFs_BHNS-NSNS}. Both noticeably improve over the prior for $L$ and $K{\rm sym}$, but the NSNS binary gives better constraints.

Much of the benefit of measuring the masses and radii of multiple NSs comes from the increased knowledge of the mass-radius relationship, the shape of which can vary significantly for different EOSs. Compared to posteriors for a single measurement, the posteriors for five frequency measurements shown in Figure~\ref{fig:combineRSFs_4data} may similarly benefit from having information about the mass-frequency relationship, but this systematic benefit is combined with statistical improvements from having more data.
To separate these effects, we compare posteriors of an inference using two frequency measurements from NSs with very different masses -- which maximises the influence of the (near-linear) mass-frequency relationship -- to posteriors of inferences using two identical measurements.
From this Figure however we can see that there is little benefit to probing the mass-frequency relationship, as two frequency measurements from identical NSs can be just as good for constraining the symmetry energy parameters as two measurements from NSs with very different masses.
Any multimessenger RSFs and GW events detected after the first will therefore be useful for constraining the symmetry energy, not just those that probe a new section of the mass-frequency relationship.

\begin{figure}
\centering
\includegraphics[width=0.45\textwidth,angle=0]{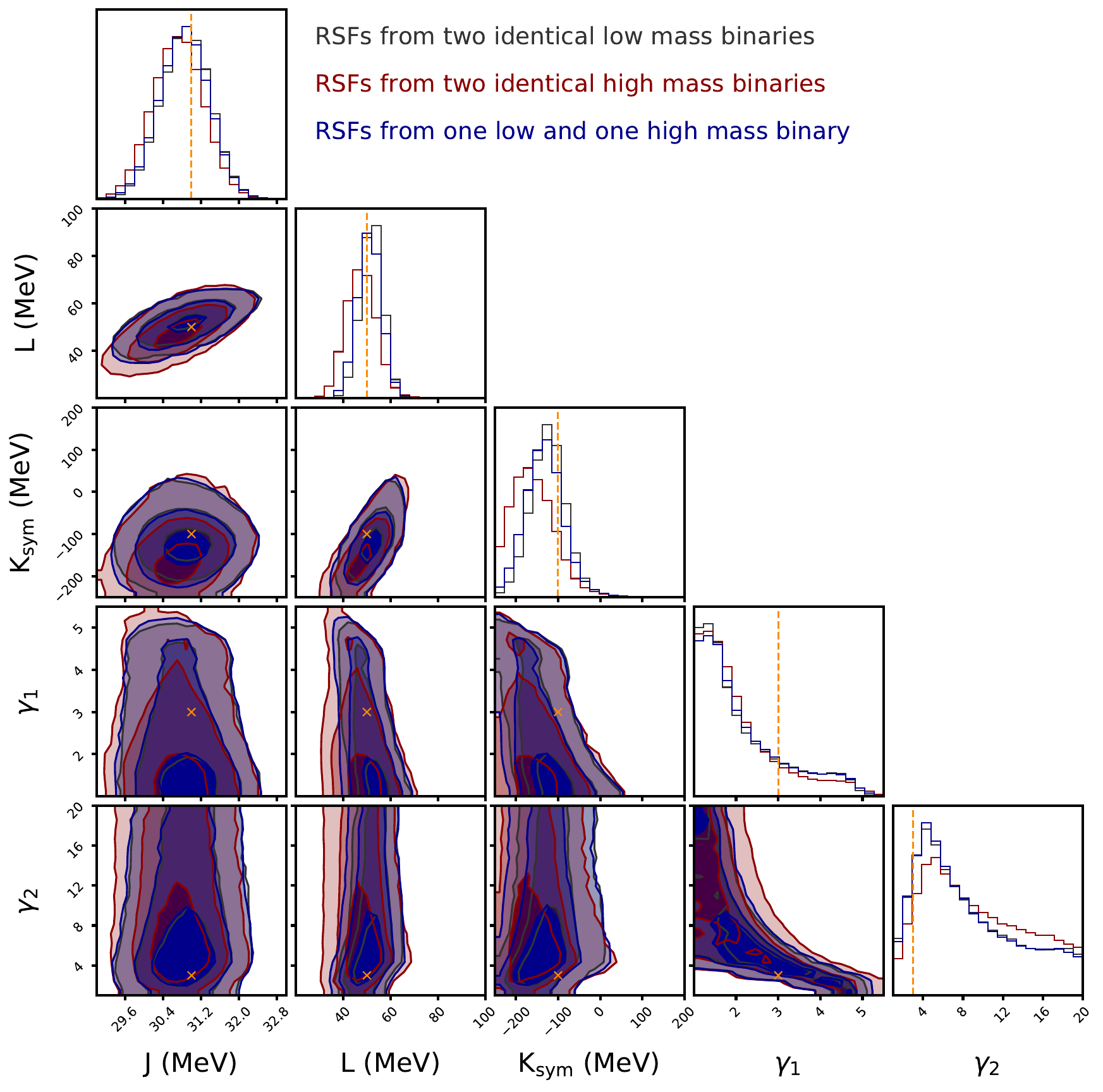}
\caption{Similar to Figure~\ref{fig:combineRSFs_4data}, but isolating the benefit of probing the mass-frequency relationship, which is shown to be negligible as the blue posteriors do not recover the injected parameters significantly better than the other posteriors. Each inference uses $i$-mode frequency measurements from two RSFs, with grey (red) using two identical low mass (high mass) NS progenitors, while blue uses one low mass and one high mass progenitor. All three require a stable EOS and reasonable maximum mass, and include various nuclear data.} 
\label{fig:combineRSFs_extremes}
\end{figure}

\section{Discussion}\label{sec:discuss}

From Figures~\ref{fig:Muncertainty},~\ref{fig:combineRSFs_4data},~\ref{fig:combineRSFs_BHNS-NSNS} and~\ref{fig:combineRSFs_extremes} we can see that there is some benefit to having measurements of the $i$-mode frequency from multiple different multimessenger events, and that the mass of the NS which produced a RSF has little impact on the symmetry energy constraints that can be obtained from it. This does not mean that NS mass is irrelevant for RSFs, as it may still have an effect on the RSF emission mechanism that could make it easier to detect RSFs from higher or lower mass NSs, but once a RSF is detected alongside GWs it is as useful as any other.

Figure~\ref{fig:combineRSFs_4data} also shows that there is little change in the inferred values of the core polytrope parameters when using additional $i$-mode frequency measurements, which is due to the $i$-mode frequency having little dependence on the NS core \citep[as found in][]{Neill2023Constraining}. This sets it apart from bulk properties such as NS radius and tidal deformability, and means that it can be used to probe the composition of the NS crust and outer core without concern for the type of matter present deeper within the core, which is currently uncertain.

Higher-order terms in the symmetry energy expansion control the EOS further away from saturation density. In our model the expansion coefficients of the symmetry energy at higher order than $K_{\rm sym}$ are correlated with the lower order parameters, but it has been shown that the size of the effects of the fourth-order term $Q_{\rm sym}$ may not be negligible down to 0.5$n_{\rm sat}$ and up to 1.5$n_{\rm sat}$, with the crust composition showing some sensitivity to that parameter \citep{Carreau2019Bayesian}. It is therefore important to extend the model to include freedom in that parameter in future studies, as degeneracies between it and the $J$, $L$ and $K_{\rm sym}$ parameters considered in this work may affect their posteriors.

\begin{figure}
    \centering
    \includegraphics[width=0.45\textwidth,angle=0]{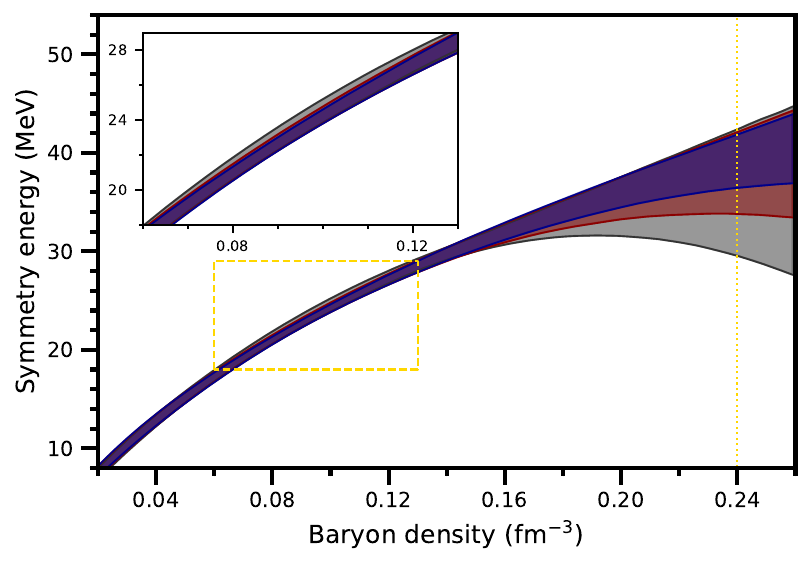}
    \caption{Nuclear symmetry energy posteriors corresponding to the posteriors for its parameters shown in Figure~\ref{fig:combineRSFs_4data}, showing how they are related within our model-dependent setup. At any given density the shaded regions contain the central 90\% of symmetry energy values for samples in the posteriors, and the same colours are used as in Figure~\ref{fig:combineRSFs_4data}. The inset highlights the density range that $i$-mode frequency measurements inform us of (indicated by the yellow box), which is around the crust-core transition density. The vertical line indicates the density beyond which the symmetry energy has no effect on our models since we switch to a polytropic model for the NS EOS. As we have seen in Figures~\ref{fig:Muncertainty}-\ref{fig:combineRSFs_extremes}, the nuclear data constrains the magnitude of the symmetry energy at a density of around 0.1 fm$^{-3}$. Resonant shattering flares meanwhile constrain the slope and curvature of the symmetry energy at that density: differences that show up at higher densities where the improvement from the astrophysical data is more apparent.}
    \label{fig:Esym_nb}
\end{figure}

The symmetry energy expansion is around nuclear saturation, but the $i$-mode frequency is sensitive to neutron star properties at the crust core transition around half that density. Model dependencies arise in our posteriors for the parameters of the expansion from how the symmetry energy may evolve between these densities in our model. Therefore, while the symmetry energy parameters at saturation are more commonly considered in the literature, it may be more useful to examine posteriors on the symmetry energy itself. Figure~\ref{fig:Esym_nb} shows posteriors for the nuclear symmetry energy that, for our meta-model, correspond to the posteriors for its first three parameters shown in Figure~\ref{fig:combineRSFs_4data}, incorporating effects of the higher-order parameters that are not free within our model. Figure~\ref{fig:Esym_nb} therefore shows what the symmetry energy parameter posteriors actually mean in the context of our meta-model, and we see that $i$-mode frequency measurements mainly inform us of the symmetry energy around half saturation density (highlighted in the inset). Freedom in how the symmetry energy varies with density for any particular sample is also limited by model-dependencies, but that is not shown in here.

One should interpret the symmetry energy constraints above nuclear saturation density in Figure~\ref{fig:Esym_nb} with care. Although they appear improved with $i$-mode measurements, the lack of freedom for the terms at higher order than $K_{\rm sym}$ in our model reduces the freedom to explore density dependencies at higher densities and leading to an over-optimistic constraint there. The density to which the constraints from our model are robust should be explored further. However, the constraints up to saturation density are on a firmer footing, as that is where the $i$-mode is sensitive to.

The NSNS binary used for Figure~\ref{fig:combineRSFs_BHNS-NSNS} had higher mass uncertainty and lower frequency uncertainty than the BHNS binary, so its better parameter recovery indicates that the frequency uncertainty has a more significant effect on these posteriors. The frequency uncertainty from equation~\eqref{eq:freq_spread} is lower for the NSNS binary because of its lower chirp mass, so applying this result more broadly would suggest that binaries with lower masses are slightly better for RSF and GW coincident timing. The difference between the posteriors for two high and two low mass NSNS binaries in Figure~\ref{fig:combineRSFs_extremes} appears to support this. However, as equation~\eqref{eq:freq_spread} is only a rough approximation, binary mass might not actually be a primary contributor to RSF duration, and variables unrelated to the binary -- such as the configurations of instruments involved in a detection -- could be more important for the uncertainty in the duration of real multimessenger events.

The mass-frequency relationships shown in Figure~\ref{fig:mfrt_vary1} are all very similar and near-linear, meaning that when we measure mass and frequency we are effectively just constraining the intercept of a linear relationship. Measurements from different NSs therefore all effectively inform us of the same one property. This is supported by Figure~\ref{fig:combineRSFs_extremes}, which shows that constraining the mass-frequency relationship does not provide much advantage over measuring a single point along it.
The changes we see in Figure~\ref{fig:combineRSFs_4data} when adding more multimessenger measurements are therefore primarily due to the statistical improvements of measuring the same property multiple times. It is possible that the similarity of the mass-frequency relationships shown in Figure~\ref{fig:mfrt_vary1} is particular to our meta-model and that methods of generating NS models that have more freedom could allow for more variety, in which case it would be more beneficial to detect RSFs from NSs with a variety of different masses.

While in this work we have not carried forward any information about tidal deformability from GW analysis to the meta-model parameter inference, combining tidal deformability inferences with $i$-mode frequency measurements could be extremely interesting, as the former is sensitive to the properties of whichever type of matter exists in the NS core while the latter depends on the nucleonic crust. 
An $i$-mode frequency measurement from a multimessenger RSF and GW event could be used to confidently constrain the properties of nucleonic matter in NSs at a low density, which could then be expanded up to higher densities to compare to the properties inferred from tidal deformability. Inconsistency between the properties of nucleonic matter inferred at high and low densities would indicate that matter is not nucleonic in the core or that current nucleonic models are insufficient to explain both high and low density matter, giving some insight into the mysterious nature of extremely dense nuclear matter.

All of the posteriors we have shown are for inferences which included the experimental nuclear data listed in Table~\ref{tab:nuclear}. Without this data to reduce the size of the symmetry energy parameter space subsequent frequency measurements are much less useful, as degeneracies between the different parameters would be much more significant. It is therefore important to not just analyse this astrophysical data in a vacuum, but rather combine it with terrestrial nuclear physics.

\subsection{Model dependence and caveats}

The Skyrme model is one model for the nuclear interaction and has different density dependencies than, for example, relativistic mean-field models. However, extended Skyrme models increase the freedom in the density dependence and offset this problem. Nevertheless, the possibility remains that systematically different predictions of the EOS and crust composition may be made by different nuclear models, and should be explored in the future.

Our inferences using \texttt{Bilby} in Section~\ref{sec:bilby} contain several simplifications to make them quicker.
For example, while the tidal deformability prior we used in Section~\ref{sec:bilby} was roughly based on our meta-model it did not exactly follow it, and in particular we did not account for how the tidal deformabilities of two NSs with the same EOS and different masses should be correlated. Inferring the meta-model parameters and then calculating the tidal deformability would be a more complete way to account for our choice of model, or alternatively a different tidal deformability prior could be used that considers a wider variety of NS models than just those produced by our meta-model. We also fixed several extrinsic and intrinsic properties, such as the sky position and luminosity distance of the binary, and for real data these would be included, making these inferences significantly longer and possibly increasing NS mass uncertainty through correlations with these extra parameters. However, even if mass uncertainty were significantly higher, from Figure~\ref{fig:Muncertainty} it does not seem likely that it would strongly affect the symmetry energy parameter inferences.

It is worth noting that for a real multimessenger event it would be better to perform a single parameter inference incorporating both the GW signal and coincident timing, rather than separating them into two as we have done in this work. It would allow for complementary information from these data to be combined in a statistically consistent way, and help to avoid losing relevant information, such as how we did not carry forward the tidal deformability posteriors from \texttt{Bilby} to our meta-model parameter inferences. In this work it has not been a significant problem and separating it into two steps made it our inferences simpler to perform, but when working with real data we should be more cautious.

We chose to construct our prior for the meta-model parameters to be uniform in the viable $J-L-K_{\rm sym}-\gamma_1-\gamma_2$ parameter space. However, as we have constructed the core with polytropes that cause the sound speed to monotonically increase with density, fixed the density of the transition to the polytropic model (to $1.5n_{\rm sat}$), and required that viable EOSs remain causal at least until the central density of their most massive NSs, this prior will be somewhat biased towards combinations of $J$, $L$ and $K{\rm sym}$ that result in lower sound speeds at the polytropic transition density, since lower sound speeds allow wider viable spaces for $\gamma_1$ and $\gamma_2$. This is not necessarily physical, as other models -- such as those that include exotic matter in the NS core -- can have sound speed vary more freely throughout the core and so do not need to have low values at low densities in order to remain causal. 

\begin{figure}
    \centering
    \includegraphics[width=0.45\textwidth,angle=0]{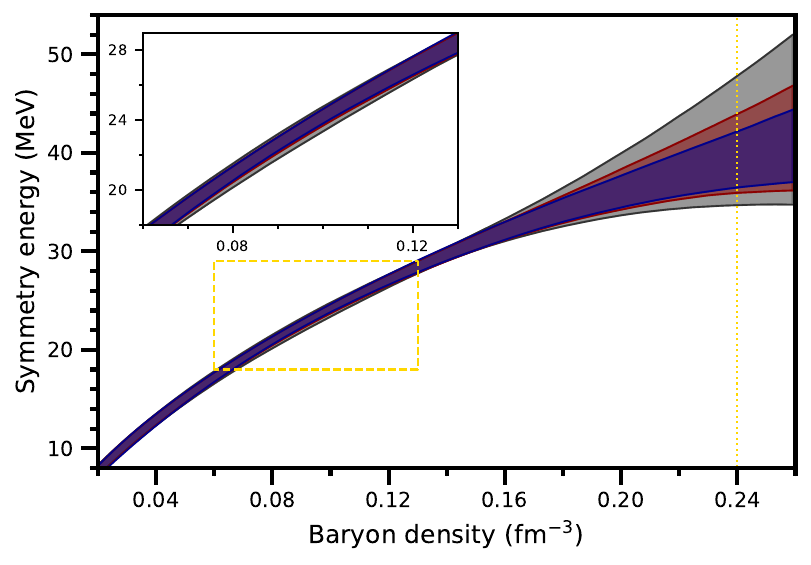}
    \caption{Similar to Figure~\ref{fig:Esym_nb}, but for posteriors obtained when using a prior designed to reduce the impact of how we chose to construct models for the NS core. The posteriors informed by several $i$-mode frequency measurements are similar to those in Figure~\ref{fig:Esym_nb}, indicating that while the choice of prior has some influence it is not very important when considering a reasonable amount of data.}
    \label{fig:Esym_nb_flatJLKprior}
\end{figure}

Figure~\ref{fig:Esym_nb_flatJLKprior} is similar to Figure~\ref{fig:Esym_nb} but attempts to remove the bias towards low sound speeds at the polytropic transition density by using a prior that is uniform in the 3D $J$-$L$-$K_{\rm sym}$ space obtained when marginalising over the core polytrope parameters. This is done by giving any 5D sample a prior probability that is inversely proportional to the area of the viable $\gamma_1-\gamma_2$ space at its $J$-$L$-$K_{\rm sym}$ values. This causes the prior to artificially favour NS EOSs that are soft above $1.5n_{\rm sat}$, rather than soft below $1.5n_{\rm sat}$ as the prior used in the rest of this work does. However, by comparing Figures~\ref{fig:Esym_nb_flatJLKprior} and~\ref{fig:Esym_nb} we can see that the posteriors informed by $i$-mode frequency data are similar regardless of the choice of prior, indicating that it is not of great importance. This means that, so long as a reasonable prior is used, the quantitatively meaningful results for real data should not be too sensitive to the choice of prior.

The frequency uncertainty obtained from coincident RSF and GW timing is significant. By comparing Figures~\ref{fig:mfrt_vary1} and~\ref{fig:fm_evidence} we can see that the uncertainty we assumed with equation~\eqref{eq:freq_spread} is not small when compared to the effects of changing the meta-model parameters, so reducing it could help to further constrain the parameters.
Equation~\eqref{eq:freq_spread} uses a rough approximation for the duration of the resonance and does not involve any consideration of the uncertain strength of the neutron star crust, and so the actual range of GW frequencies during which the $i$-mode is sufficiently excited to shatter the NS crust may be significantly longer or shorter.
However, coincident RSF and GW timing might not be the only way to infer the $i$-mode frequency from a detected RSF. Other methods involving analysis of the RSF lightcurve -- its duration, luminosity, variability timescale, etc. -- could provide additional insight into the $i$-mode and the NS crust, which could be used to improve a frequency measurement or obtain one where no GWs are detected.

\section{Conclusion}\label{sec:conclude}
The mass of a neutron star is important for its bulk properties such as radius and tidal deformability. In particular, the relationships between these properties and mass are sensitive to the equation of state of the neutron star core, so probing them by obtaining measurements from several neutrons stars with known different masses provides new insights into the physics of dense matter. In this work, we have investigated whether having multiple measurements of the crust-core interface mode's frequency -- as could be obtained from coincident timing of resonant shattering flares and gravitational waves -- would provide a similar improvement for our understanding of matter within neutron stars. Using realistic uncertainties in neutron star masses inferred with gravitational wave analysis and conservative uncertainties in $i$-mode frequency measurements obtained from coincident timing, we found that combining frequency measurements from NSs with different masses noticeably improves the nuclear symmetry energy constraints obtained from them. We found that this is not so much due to the mass-dependence of the $i$-mode frequency -- as it is only a weak dependence and is similar across the range of neutron star models we considered -- but rather is the general statistical improvements from having multiple measurements of the same property. This means that any additional frequency measurements could be useful, not just those from NSs with masses that are well-determined or that have not been seen previously.

While obtaining multiple $i$-mode frequency measurements from coincident RSF and GW timing could strengthen constraints on the NS model and the nuclear symmetry energy parameters, it is just as important to combine these constraints with those from other sources to break degeneracies between the symmetry energy parameters. 
Bulk NS properties -- which are more commonly inferred from NS observations -- might not be reliable for this as the nature of the NS core is unknown, but other NS asteroseismic observables and data from nuclear experiment can fill this role, as NS meta-models such as the one used in this work allow us to consistently connect neutron star structure to the fundamental properties of nucleonic matter. In light of the results of the recent PREX-II \citep{Adhikari2021PREXII} and CREX \citep{Adhikari2022CREX} experiments, combining data from new methods of probing nuclear matter is particularly interesting as it opens new ways to explore the tension between them.

Efforts to probe NSs with astrophysical observables are progressing at a steady pace, with constraints on masses, radii and tidal deformabilities \citep[e.g.][]{abbott2017gw170817,abbott2019properties,riley2019NICER,riley2021NICER,Fonseca2021Refined} giving us an increasing amount of insight into their structure. While a multimessenger RSF and GW event has yet to be observed, the findings of \citet{Neill2023Constraining} -- that coincident timing of these events allows us to probe a different region of the NS to most observables -- and of this work -- that even with realistic uncertainties we only require a single such detection to obtain strong constraints -- makes them highly interesting. However, this work has also shown that observing several multimessenger events is beneficial for our understanding of nuclear matter, particularly when combined with data from experimental nuclear physics.

Multimessenger detections of binary mergers are of great interest to the astrophysical community, and we remain optimistic about the possibility of observing multimessenger RSF and GW events in the remainder of the O4 LIGO/Virgo observing run, or early in O5. We have shown such observations with the sensitivity of those runs would allow us to obtain strong astrophysical constraints on NS crust composition and the nuclear symmetry energy, meaning that they are interesting for nuclear physics as well.

\section*{Acknowledgements}
This work used the Isambard 2 UK National Tier-2 HPC Service (http://gw4.ac.uk/isambard/) operated by GW4 and the UK Met Office, and funded by EPSRC (EP/T022078/1). The authors gratefully acknowledge the University of Bath’s Research Computing Group (doi.org/10.15125/b6cd-s854) for their support in this work. DN and DT were supported by the UK Science and Technology Facilities Council (ST/X001067/1) and the Royal Society (RGS/R1/231499). WGN was supported by NASA (80NSSC18K1019). We also thank Natalie Williams and Hannah Middleton for useful discussion about the usage of \texttt{Bilby}.

\section*{Data Availability}
The posterior samples generated for this work are available upon request. This work made use of \texttt{emcee}~\citep{foreman-mackey2013emcee} version 3.1 and \texttt{Bilby}~\citep{Ashton2019BILBY} version 2.1.

\bibliography{RSFSymmetry}{}
\bibliographystyle{mnras}

\bsp	
\label{lastpage}
\end{document}